\documentclass[pss]{wiley2sp} 
\usepackage{color}
\usepackage{amsmath}
\usepackage{bm}              
\usepackage{w-greek}         

\newcommand{\ee}{\mathrm{e}}
\newcommand{\hh}{\mathrm{h}}

\begin{document}

\title{Electron confinement in graphene  with gate-defined quantum dots}

\titlerunning{Short title }

\author{%
  Holger Fehske\textsuperscript{\Ast,\textsf{\bfseries 1}},
  Georg Hager\textsuperscript{\textsf{\bfseries 2}},
  Andreas Pieper\textsuperscript{\textsf{\bfseries 3}}}

\authorrunning{First author et al.}

\mail{e-mail
  \textsf{fehske@physik.uni-greifswald.de}, Phone:
  +49-3834-864760, Fax: +49-3834-864701}

\institute{%
  \textsuperscript{1}\,Institut f\"ur Physik, Ernst-Moritz-Arndt-Universit\"at
  Greifswald, 17487 Greifswald, Germany\\
  \textsuperscript{2}\,Regionales Rechenzentrum Erlangen, 
Universit\"at Erlangen-N\"urnberg, 91058 Erlangen, Germany\\
  \textsuperscript{3}\,Institut f\"ur Physik, Ernst-Moritz-Arndt-Universit\"at
  Greifswald, 17487 Greifswald, Germany}

\received{XXXX, revised XXXX, accepted XXXX} 
\published{XXXX} 

\abstract{%
%
%
%
\abstcol{%
We theoretically analyse the possibility to electrostatically confine electrons in circular quantum dot arrays, impressed on contacted graphene nanoribbons by top gates. Utilising exact numerical techniques, we compute the scattering efficiency of a single dot and demonstrate that 
for  small-sized scatterers the cross-sections are dominated by quantum effects, where resonant scattering leads to a series of quasi-bound dot states. Calculating the conductance and the local density of states for quantum dot superlattices we show that the resonant carrier transport through such graphene-based nanostructures  
 can be easily  tuned by varying the gate voltage.   
}{%
   }}
\titlefigure[height=3.cm]{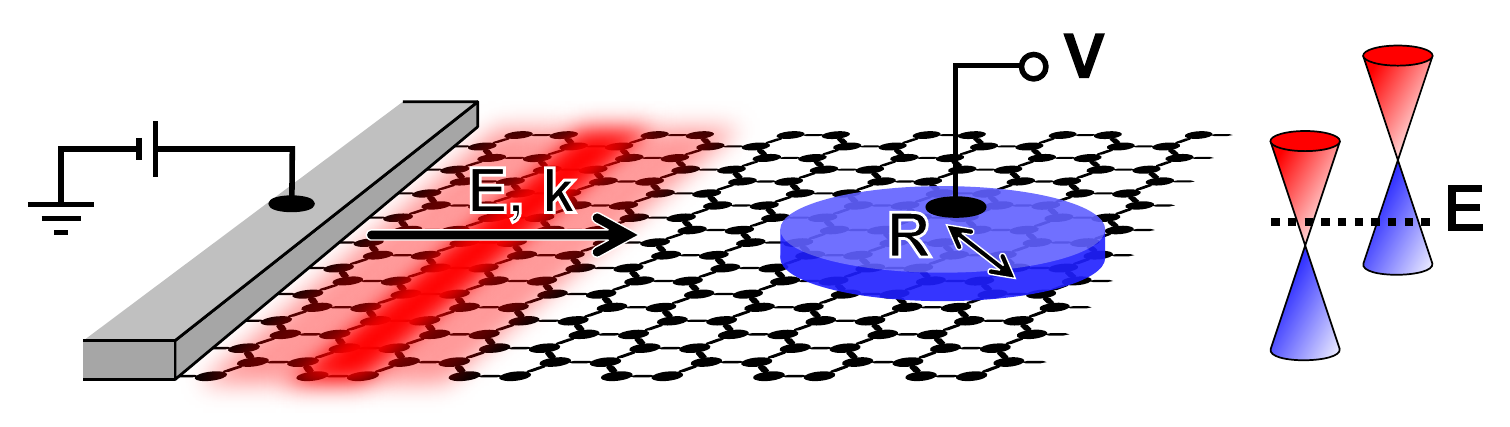}
\titlefigurecaption{%
  Schematic representation of a Dirac electron wave packet impinging on a circular, electrostatically defined quantum dot.}

\maketitle   
\section{Introduction} High-quality graphene nanostructures are the most likely building blocks in future electronics, plasmonics and photonics assemblies.
Most of their striking features arise from the strictly two-dimensional, honeycomb lattice structure of the basic material, which causes the nontrivial topology of the electronic wave function and an almost linear low-energy spectrum of the chiral quasiparticles (charge carriers) near the so-called Dirac nodal points~\cite{CGPNG09}.  From an application-technological point of view, the tunability of the transport properties of graphene by external electric and magnetic fields is of particular importance~\cite{Go11}. This allows a controlled modification of selected areas of the sample by gating.  For instance, applying nanoscale top gates on  graphene nanoribbons (GNRs) or bilayer graphene, single or double quantum dots  with particular shape have been produced in recent experiments~\cite{LHV10,AMY12,MKHWPS14}.  

For a large circular dots, the refraction at the boundary  gives rise to two coalescing 
caustics that focus the electron density in the disk-shaped  region~\cite{CPP07,AU13}.  Resonances in the conductance~\cite{BTB09}  and the scattering cross section~\cite{HBF13a}  indicate electron confinement for small quantum dots  in monolayer graphene as well. Apparently the difficulty to localise Dirac electrons by electrostatic potential barriers (which is related to the Klein tunnelling phenomenon) could be overcome: Because of the circular symmetry,  the angle of incidence at the dot's boundary is a constant of motion so that electrons with nonzero angular momentum are confined~\cite{BTB09}. Forward scattering and Klein tunnelling will also be suppressed 
if a Fano resonance arises between the background partition and the resonant contribution to electron scattering~\cite{HBF13a}. These results, obtained within continuum Dirac theory, were numerically confirmed for a lattice model~\cite{PAS11,PHF13}.  The role of resonances have been analysed in various graphene scattering experiments~\cite{HZZ09,MOWBFGGBFM11}.  
Note that the scattering (transport and optics) of massless Dirac fermions on potential barriers substantially differs from that of massive chiral Dirac fermions or Schr\"odinger electrons~\cite{AFkt11}.

\section{Model and numerical methods} In this work,  we address the scattering and (linear) transport in armchair  GNRs with gate-defined quantum dots in the end-contacted lead-sample geometry most relevant for experiments~\cite{KR12,RG13,LRG13}. The leads can be viewed as semi-infinite quantum wires; the contacts are metallic or gated graphene. 

We describe the sample by the tight-binding Hamiltonian 
\begin{equation}
     {H} =  \sum_{i}V_i^{} {c}_i^{\dag} {c}_i^{} 
           -t \sum_{\langle ij \rangle}({c}_i^{\dag} {c}_j^{} + \text{H.c.})\,,
\label{eq:H}
\end{equation}
where $c_i^{(\dag)}$ annihilates (creates) an electron  at Wannier site $i$ of a honeycomb lattice with carbon-carbon distance $a=1.42$~\AA. Here the hopping amplitude between nearest neighbour sites $\langle ij\rangle$ is given by $t$ ($\simeq$ 3eV) and the site-dependent 
 potentials $V_i=V \sum_{n} \Theta (R-|\vec{r}_i-\vec{r}_{n}|)$ 
model electrostatically defined quantum dots with radius $R$  centred at positions $\vec{r}_n$.

In the limit of vanishing bias voltage, the conductance between the left (L) and right (R) leads can be obtained within the limits of the Landauer-B\"uttiker approach: 
\begin{equation}
G_{\rm LR}= \frac{\ee^2}{\hh} \sum_{n \in L, m \in R} | S_{n,m} |^2\,,
\label{eq:cond}
\end{equation}
where  $S_{n,m}$ is the scattering matrix between all open (i.e., active) lead channels~\cite{Dat95}. We furthermore exploit the local density 
of states (LDOS) at a certain site $i$ of the sample, 
\begin{equation}
\text{LDOS} (E) = \sum_{l} |\langle i | l\rangle |^2 \delta (E-E_l)\,,
\label{eq:ldos}
\end{equation}
as a probe of a possible electrostatic confinement
of the charge carrier~\cite{PHWF14,SB14}.
In Eq.~(\ref{eq:ldos}), $E$ is the particle's energy
and the sum extends over all single-electron eigenstates
$|l\rangle=c_l^\dagger |0\rangle$ of $H$ with energy $E_l$. We use the kernel polynomial method~\cite{WWAF06} and the `Kwant' software package~\cite{Kwant14}
for evaluating the conductance and LDOS numerically. 
\begin{figure}[t]
\includegraphics[width=.91\linewidth]{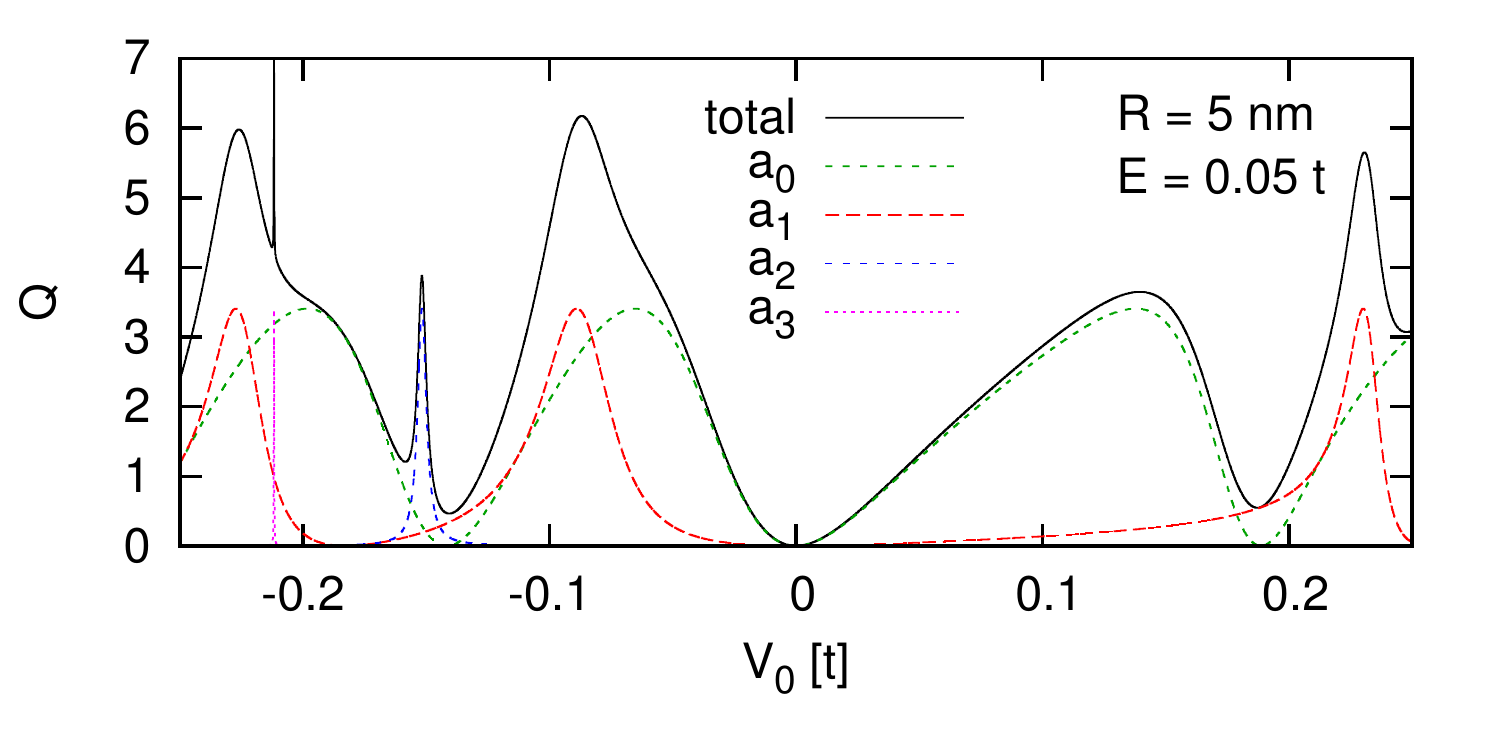}\\
\includegraphics[width=.95\linewidth]{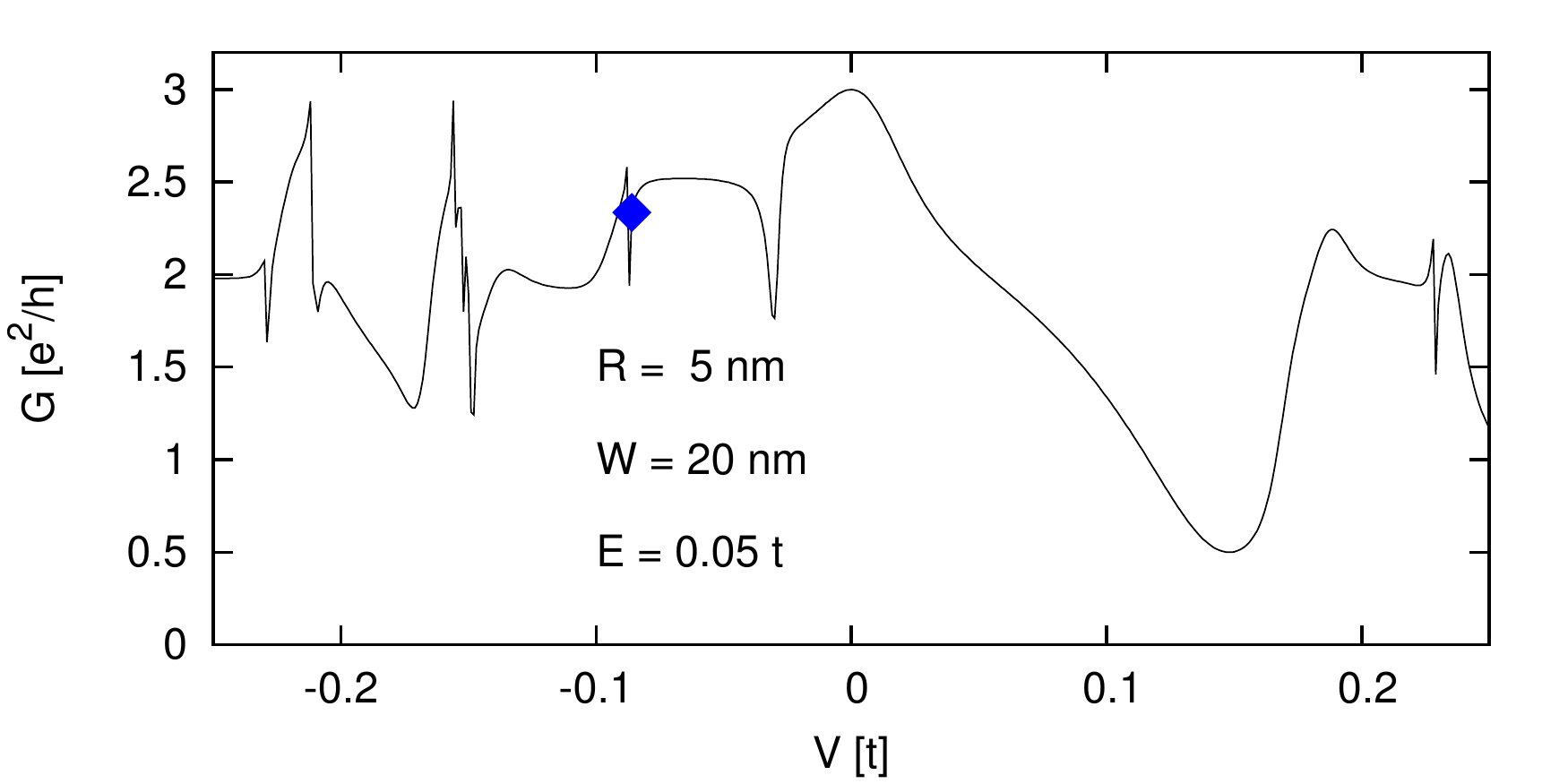}\\
\includegraphics[width=\linewidth]{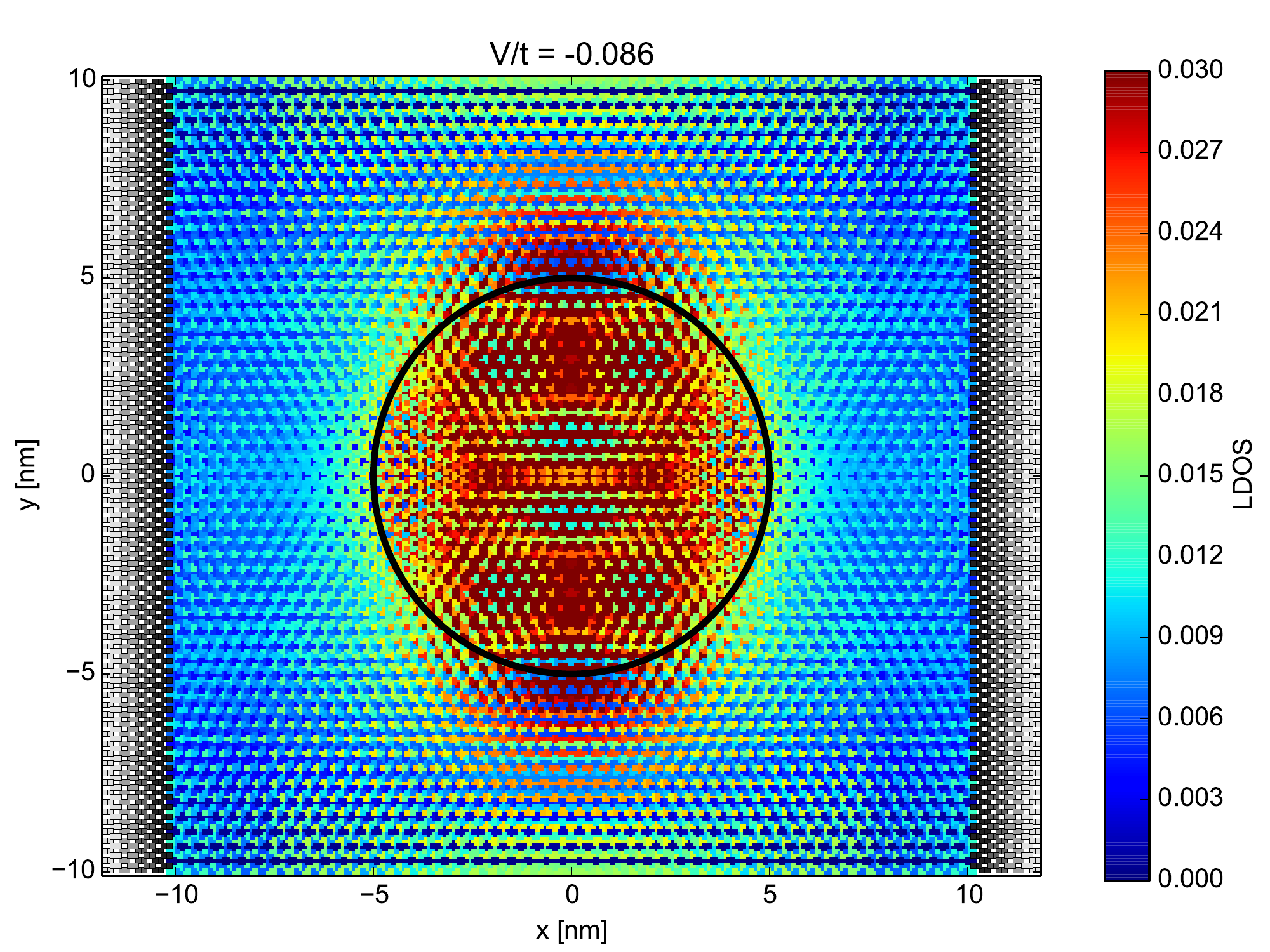}
\caption{Total and partial scattering efficiencies  (top), conductance (middle) and LDOS (bottom) of a single quantum dot 
(the black circle  specifies the dimension of the dot).   For further explanation see text.}
\label{fig1}
 \end{figure} 
\section{Results and discussion}
\subsection{Single quantum dot}
We first consider a single dot, see Fig.~\ref{fig1}.
To begin with we determine  the scattering efficiency $Q$, which is 
the scattering cross section divided by the geometric cross section,
of a plane Dirac electron wave hitting on a dot embedded 
in an infinite graphene sheet, within the Dirac approximation
(for details of such a  calculation cf. 
Ref.~\cite{HBF13a}). Since the energy is conserved in our setup 
(i.e., there is no coupling to bath degrees of freedom etc.), the scattering is always elastic (coherent). 
The data displayed in the upper panel for $E=0.05 \,t$ and $R=5$~nm
shows that a series of resonances appears when the gate voltage is varied (besides the rather broad $a_0$ and $a_1$ modes, the 
$a_{m>1}$ resonances are very sharp),
which can be attributed to the excitations of higher dot normal modes~\cite{HBF13a}. 
So each maximum marks a combination of $E$, $R$, and $V$,
where a certain normal mode fits particularly well into the dot,
while all other  modes are suppressed.
The middle panel indicates that these signatures also show up in the conductance,
which this time, however, was computed for a 20~nm wide armchair GNR placed between leads.
That is why we can assign the signal at $V\simeq - 0.09 t$ to the $a_1$ mode.
The value for $G[V=0]$  tells us that the system exhibits three open transport channels.
The lowermost panel presents the LDOS close to the $a_1$ resonance marked
in the $G$~vs.~$V$ plot by the blue diamond.
Obviously the electron is almost completely localised at the quantum dot,
where the superposition of the states with $j=\pm(m+1/2)$ leads to
the characteristic vortex pattern of the $a_1$ mode. There are six vortices close to the boundary of the dot which are reflected by three preferred 
scattering directions in the far field~\cite{HBF13a}.  
\begin{figure}[t]
\centering
\includegraphics[width=0.8\linewidth]{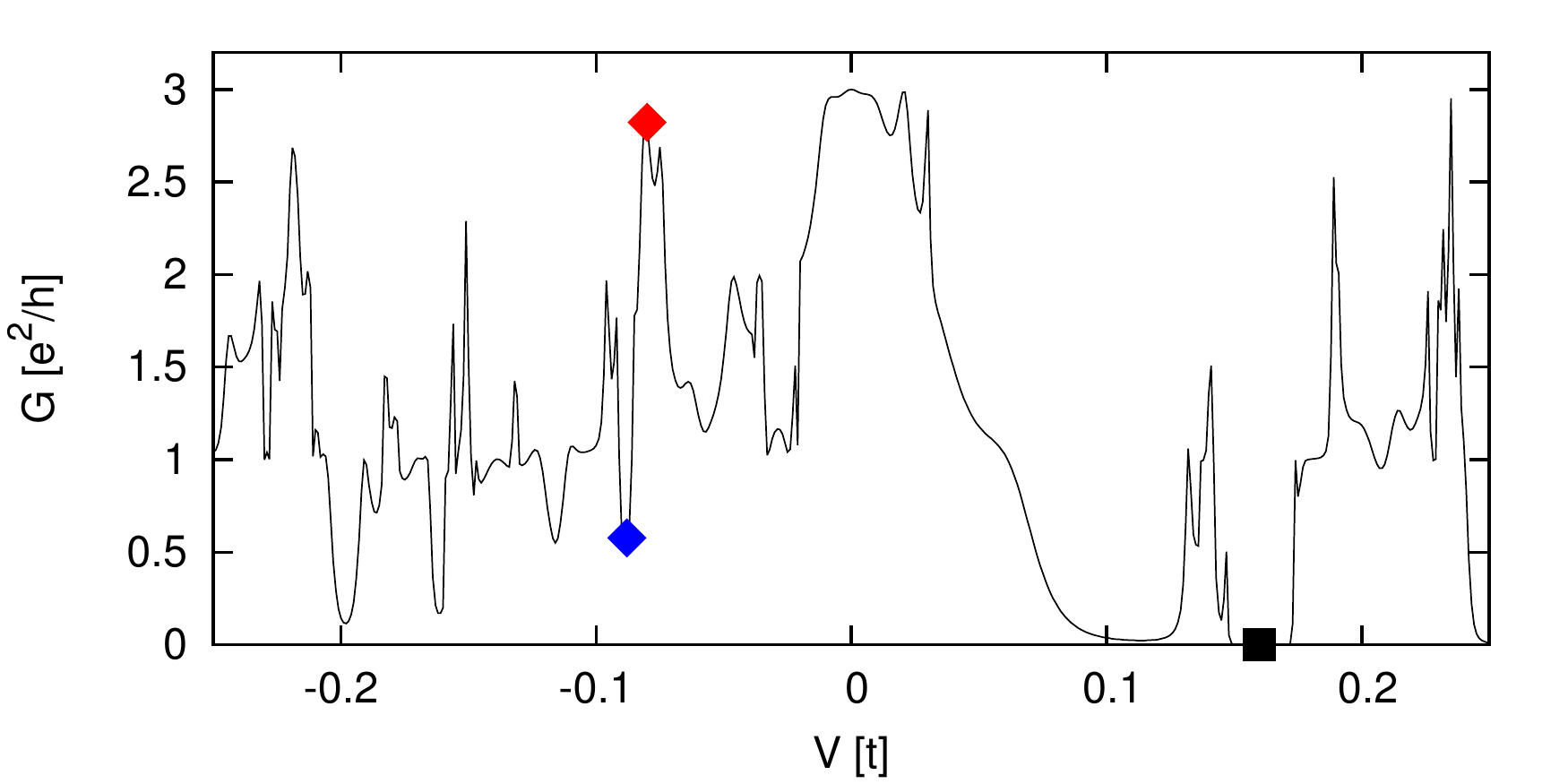}
\includegraphics[width=\linewidth]{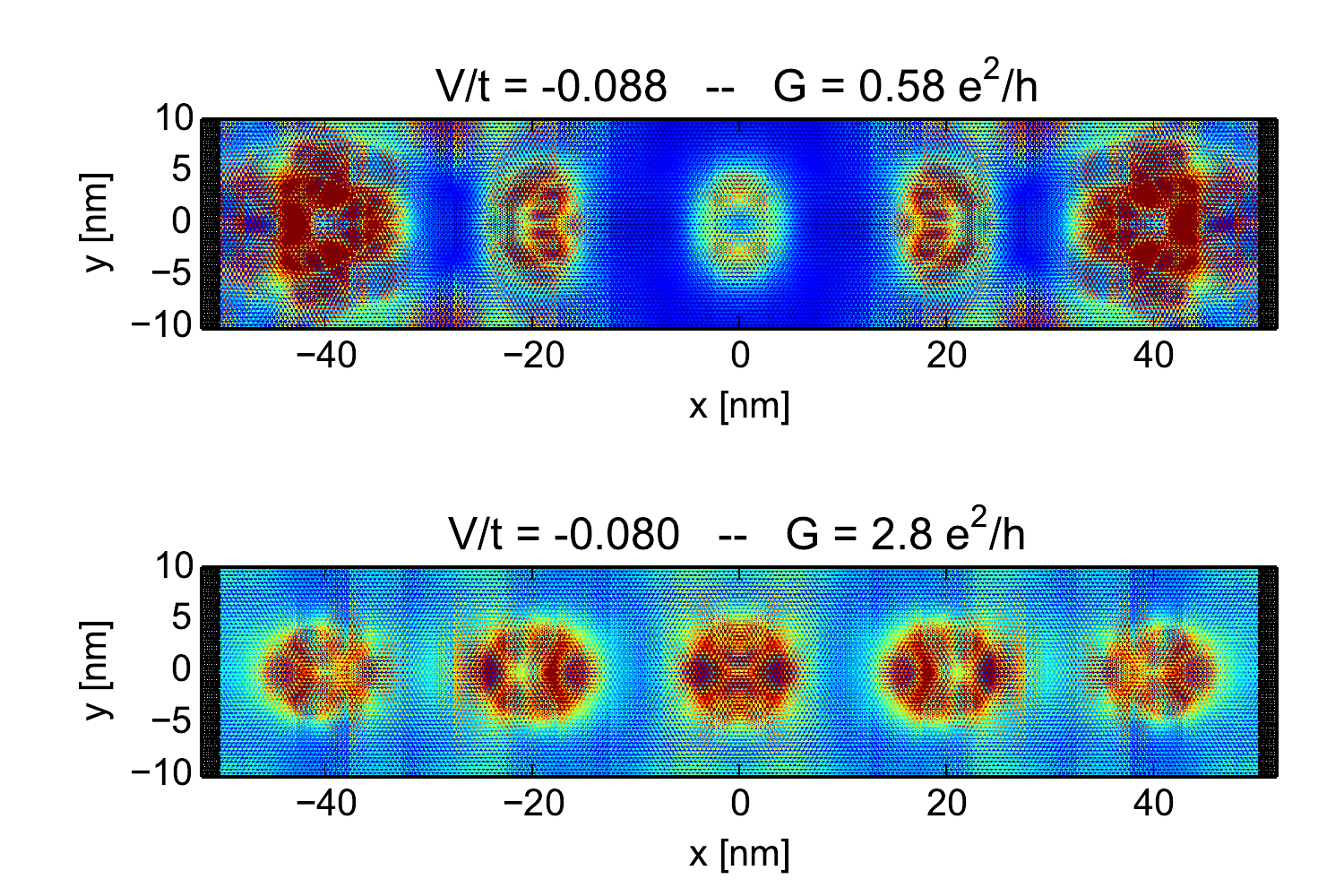}
\caption{Conductance (top) and LDOS (lower panels) of a GNR with a linear array of quantum dots. The colour code is the same as in Fig.~\ref{fig1}.}
\label{fig2}
 \end{figure}

\subsection{Qunatum dot chain}
We now study quantum dots arranged like a string of pearls.
Figure~\ref{fig2}  presents the conductance and the LDOS 
for a contacted GNR sample with five dots at intervals of 20~nm. 
Again we set $R=5$~nm, $E=0.05 t$, and the width of the armchair GNR is 20~nm. 
Note that the conductance undergoes a dramatic change 
if $V$ is slightly varied near the resonance points (see red and blue diamonds).
In this way the system may act as a switch.
The results for the LDOS provide further insight into
how the system behaves locally for  $V=-0.088\,t$ ($V=-0.080\,t$),
i.e., at the local minimum (maximum) of the conductance.
At $V=-0.088\,t$, $E=0.05 t$ (middle panel), the quantum dots are off-resonance from an eigenmode
of the particle: As a result the scattering is enhanced and the conductance is 
low (note that the figure shows a stationary situation). 
For an  applied bias $V=-0.080\,t$ (lower panel) all dots largely support the $a_1$ mode.
Since the wave functions of quasi-bound $a_1$ modes at neighbouring dots
overlap resonantly, electrons are no longer permanently confined 
to a particular dot but will be transferred through the GNR 
by an effective inter-dot hopping process. Remarkably, in this case the conductance is almost as large as 
for the GNR without quantum dots ($G[V=0]$) where the LDOS is the same (i.e., uniform) everywhere between the leads.
\begin{figure}[t]
\centering
\includegraphics[width=0.8\linewidth]{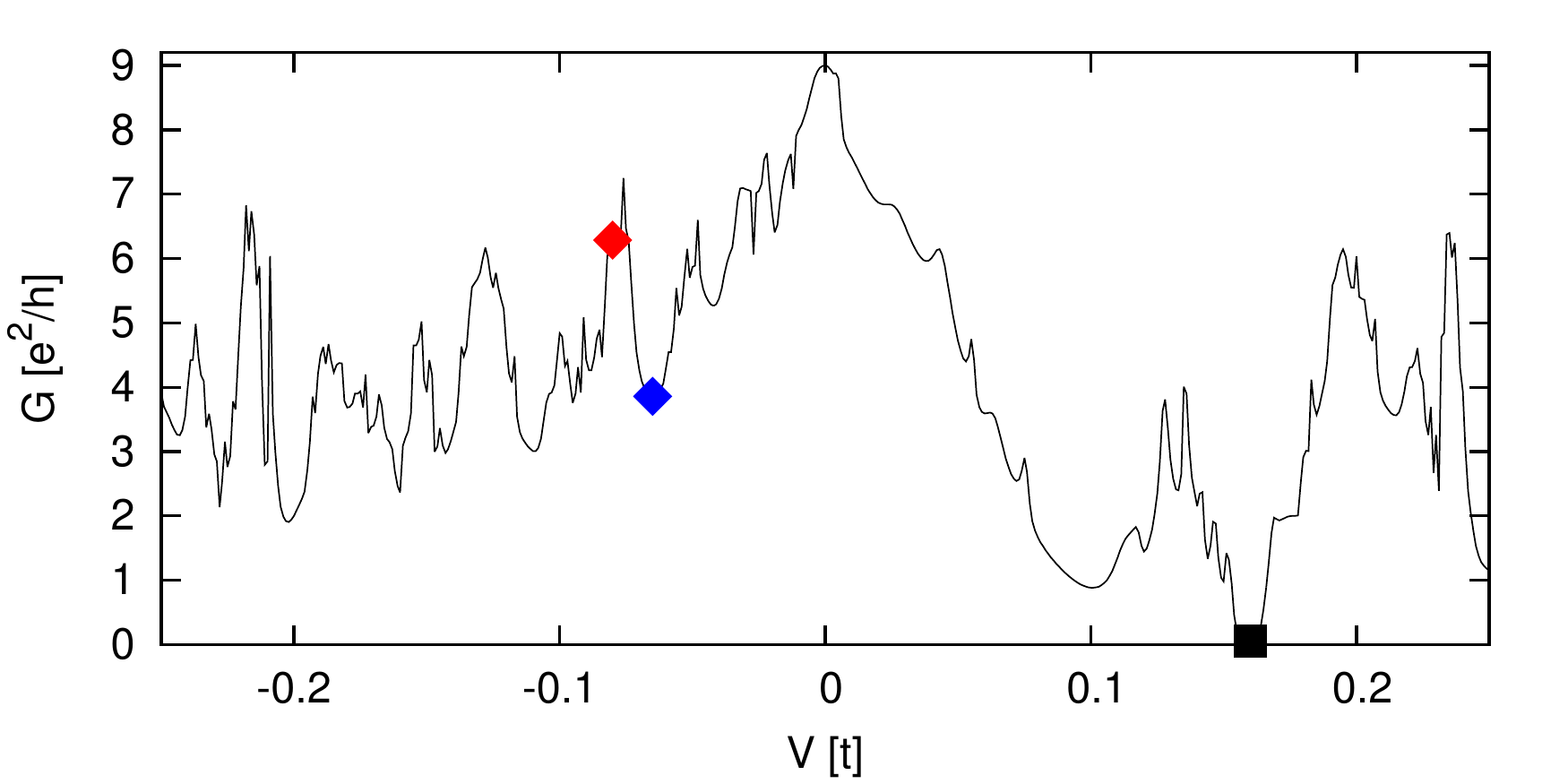}
\includegraphics[width=\linewidth]{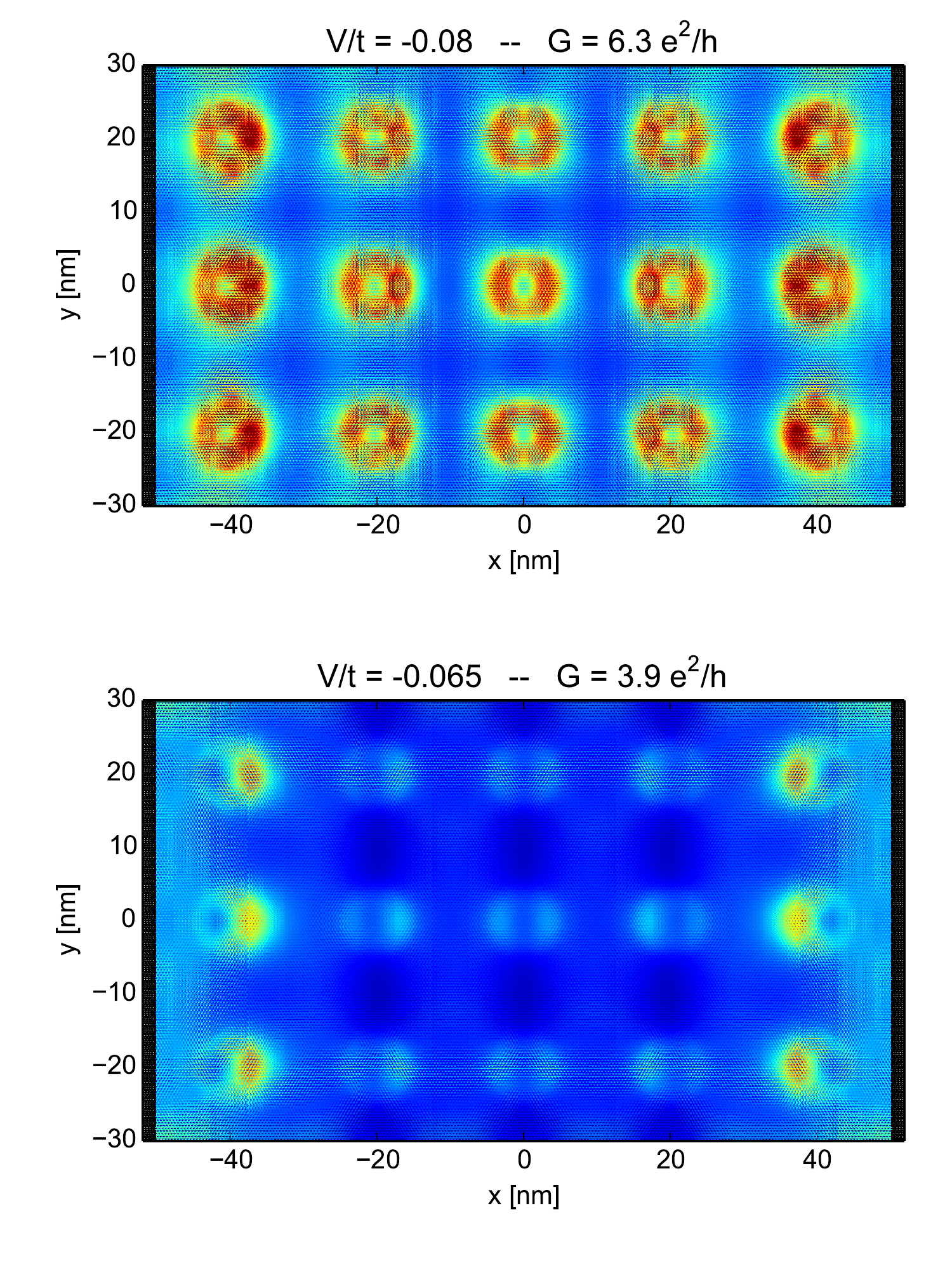}
\caption{Conductance (top) and LDOS (middle/bottom) of an armchair GNR with a square dot superlattice.}
\label{fig3}
 \end{figure} 
\subsection{Quantum dot matrix}
Figure~\ref{fig3} illustrates that the same behaviour 
could be achieved for a square quantum dot superlattice 
made on an armchair GNR (now the width of the GNR is 60~nm,
the other parameters are the same as in Fig.~\ref{fig2}). 
For $V=0$ the system has nine open transport channels. 
Again the relatively large conductance at $V=-0.080\,t$ 
(compared to $G$ for $V=-0.065\,t$) originates 
from an effective (coherent) inter-dot  transfer of the electron.
This hopping process is expected to take place on a reduced time scale~\cite{PHF13}.
As can be seen from the lowermost panel, at $V=-0.065\,t$,
the LDOS notably shrinks if we move to the centre of the GNR;
this depletion of the local particle density 
reduces the inter-dot hopping and consequently the conductance.     
\begin{figure}[t]
\centering
\includegraphics[width=\linewidth]{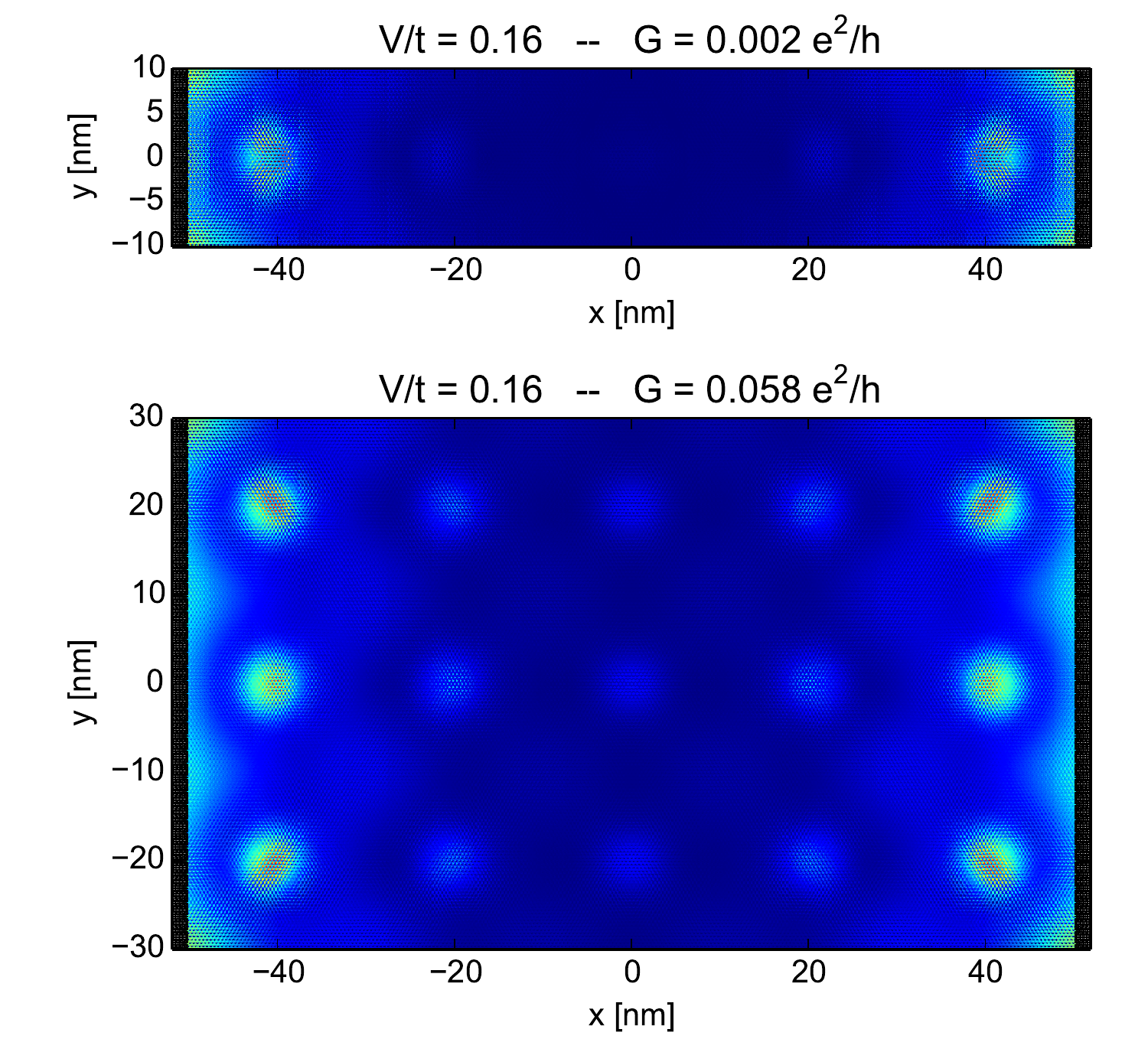}
\caption{LDOS for a linear (top) and a square (bottom)  quantum dot array on a GNR with almost vanishing conductance.}
\label{fig4}
 \end{figure} 
 
While graphene has a gapless band structure and also the armchair GNRs
considered in Figs.~\ref{fig1}-\ref{fig3} have  $G >0$  for all $E$
at $V=0$  (i.e., there are always open transport channels),
the conductance of the GNRs with imprinted quantum dots may
tend to zero in certain ranges of $V$; see the black squares in Figs.~\ref{fig2} and~\ref{fig3}.
Apparently the linear and square lattice periodic arrays of gate-defined
quantum dots give rise to new superlattice band structures,
which exhibit energy (pseudo) gaps  just as for ordinary solids~\cite{LWSYL09}. 
This is reflected in the LDOS depicted in Fig.~\ref{fig4}.

\section{Conclusions} To summarise, using exact numerics,
we have studied the electronic and transport properties 
of contacted graphene nanoribbons with gate-defined quantum dots of realistic size (i.e., 
with a few thousand carbon atoms that feel the electrostatic potential step). 
The conductivity of such systems is very sensitive to the model parameters and can even vanish. 
This provides an opportunity to use them as switches by varying the top gate potential.
A detailed analysis of the local density of states shows 
that an electrostatic electron confinement  is possible for single quantum dots
in the quantum regime due to resonant scattering. 
Thereby the incident electron is fed into vortices, which trap the particle. That means for small dots the 
particle confinement is not by total internal reflection.  In addition, for periodic quantum dot arrays a  superlattice
band structure evolves that allows for an effective inter-dot  
transport on a reduced energy scale. The phenomena detected in this work 
might be exploited in actual graphene-based devices. From a theoretical point of view, it seems worthwhile 
to study a modulation of the quantum dot barriers---via the top gates---by oscillating fields, 
which induces both inelastic tunnelling and resonant transport through excited states\cite{SHF15b}.
Another issue is the electrical and magneto transport trough quantums dots on folded graphene samples
that exhibit superlattice structures in itself~\cite{SRSH14}. We finally like to emphasize that such graphene quantum 
dots might used to construct logic gates and ``quantum computing'' devices~\cite{WL11}, which would be also 
an interesting topic for future studies. 

\begin{acknowledgement}
This work was funded by the Deutsche Forschungsgemeinschaft through 
Priority Program SPP1459 `Graphene' and the Competence Network 
for Scientific High-Performance Computing 
in Bavaria (KONWIHR III, project  PVSC-TM). 
\end{acknowledgement}

\bibliographystyle{pss}
\bibliography{ref}

\providecommand{\WileyBibTextsc}{}
\let\textsc\WileyBibTextsc
\providecommand{\othercit}{}
\providecommand{\jr}[1]{#1}
\providecommand{\etal}{~et~al.}


\begin{thebibliography}{[10]}

\bibitem{CGPNG09}
 \textsc{A.\,H. Castro~Neto},  \textsc{F.~Guinea},  \textsc{N.\,M.\,R. Peres},
  \textsc{K.\,S. Novoselov},  and  \textsc{A.\,K. Geim},
 \jr{Rev. Mod. Phys.} \textbf{81}, 109--162 (2009).


\bibitem{Go11}
 \textsc{M.\,O. Goerbig},
 \jr{Rev. Mod. Phys.} \textbf{83}, 1193 (2011).


\bibitem{LHV10}
 \textsc{X.\,L. Liu},  \textsc{D.~Hug},  and  \textsc{L.\,M.\,K. Vandersypen},
 \jr{Nano Lett.} \textbf{10}, 1623 (2010).


\bibitem{AMY12}
 \textsc{M.\,T. Allen},  \textsc{J.~Martin},  and  \textsc{A.~Yacoby},
 \jr{Nat. Commun.} \textbf{8}, 934 (2012).


\bibitem{MKHWPS14}
 \textsc{A.~M\"uller},  \textsc{B.~Kaestner},  \textsc{F.~Hohls},
  \textsc{T.~Weimann},  \textsc{K.~Pierz},  and  \textsc{H.\,W. Schumacher},
 \jr{J. Appl. Phys.} \textbf{115}, 233710 (2014).


\bibitem{CPP07}
 \textsc{J.~Cserti},  \textsc{A.~P\'alyi},  and
  \textsc{C.~P\'eterfalvi},
 \jr{Phys. Rev. Lett.} \textbf{99}, 246801 (2007).


\bibitem{AU13}
 \textsc{M.\,M. Asmar} and  \textsc{S.\,E. Ulloa},
 \jr{Phys. Rev. B} \textbf{87}, 075420 (2013).


\bibitem{BTB09}
 \textsc{J.\,H. Bardarson},  \textsc{M.~Titov},  and  \textsc{P.\,W.
  Brouwer},
 \jr{Phys. Rev. Lett.} \textbf{102}, 226803 (2009).


\bibitem{HBF13a}
 \textsc{R.\,L. Heinisch},  \textsc{F.\,X. Bronold},  and
  \textsc{H.~Fehske},
 \jr{Phys. Rev. B} \textbf{87}, 155409 (2013).


\bibitem{PAS11}
 \textsc{G.~Pal},  \textsc{W.~Apel},  and  \textsc{L.~Schweitzer},
 \jr{Phys. Rev. B} \textbf{84}, 075446 (2011).


\bibitem{PHF13}
 \textsc{A.~Pieper},  \textsc{R.~Heinisch},  and  \textsc{H.~Fehske},
 \jr{Europhys. Lett.} \textbf{104}, 47010 (2013).


\bibitem{HZZ09}
 \textsc{X.~Hong},  \textsc{K.~Zou},  and  \textsc{J.~Zhu},
 \jr{Phys. Rev. B} \textbf{80}, 241415 (2009).


\bibitem{MOWBFGGBFM11}
 \textsc{M.~Monteverde},  \textsc{C.~Ojeda-Aristizabal},  \textsc{R.~Weil},
  \textsc{K.~Bennaceur},  \textsc{M.~Ferrier},  \textsc{S.~Gu\'eron},
  \textsc{C.~Glattli},  \textsc{H.~Bouchiat},  \textsc{J.\,N. Fuchs},  and
  \textsc{D.\,L. Maslov},
 \jr{Phys. Rev. Lett.} \textbf{104}, 126801 (2010).


\bibitem{AFkt11}
 \textsc{P.\,E. Allain} and  \textsc{J.\,N. Fuchs},
 \jr{EPJB} \textbf{83}, 301 (2011).


\bibitem{KR12}
 \textsc{S.~Krompiewski},
 \jr{Nanotechnology} \textbf{23}, 135203 (2012).


\bibitem{RG13}
 \textsc{L.~Rosales} and  \textsc{J.\,W. Gonz\'{a}les},
 \jr{Nanoscale Research Letters} \textbf{8}, 5742 (2013).


\bibitem{LRG13}
 \textsc{T.~Lehmann},  \textsc{D.\,A. Ryndyk},  and
  \textsc{G.~Cuniberti},
 \jr{Phys. Rev. B} \textbf{88}, 125420 (2013).


\othercit
\bibitem{Dat95}
 \textsc{S.~Datta},
Electronic Transport in Mesoscopic Systems (Cambridge University Press,
  Cambridge, 1995).


\bibitem{PHWF14}
 \textsc{A.~Pieper},  \textsc{R.\,L. Heinisch},  \textsc{G.~Wellein},  and
  \textsc{H.~Fehske},
 \jr{Phys. Rev. B} \textbf{89}, 165121 (2014).


\bibitem{SB14}
 \textsc{M.~Schneider} and  \textsc{P.\,W. Brouwer},
 \jr{Phys. Rev. B} \textbf{89}, 205437 (2014).


\bibitem{WWAF06}
 \textsc{A.~Wei{\ss}e},  \textsc{G.~Wellein},  \textsc{A.~Alvermann},  and
  \textsc{H.~Fehske},
 \jr{Rev. Mod. Phys.} \textbf{78}, 275 (2006).


\bibitem{Kwant14}
 \textsc{C.\,W. Groth},  \textsc{M.~Wimmer},  \textsc{A.\,R. Akhmerov},  and
  \textsc{X.~Waintal},
 \jr{New Journal of Physics} \textbf{16}(6), 063065 (2014).


\bibitem{LWSYL09}
 \textsc{W.~Liu},  \textsc{Z.\,F. Wang},  \textsc{Q.\,W. Shi},
  \textsc{J.~Yang},  and  \textsc{F.~Liu},
 \jr{Phys. Rev. B} \textbf{80}, 233405 (2013).


\bibitem{SHF15b}
 \textsc{C.~Schulz},  \textsc{R.\,L. Heinisch},  and  \textsc{H.~Fehske},
 \jr{Phys. Rev. B} (2015).


\bibitem{SRSH14}
 \textsc{H.~Schmidt},  \textsc{J.\,C. Rode},  \textsc{D.~Smirnov},  and
  \textsc{R.\,J. Haug},
 \jr{Nature Communications} \textbf{5}, 5742 (2014).


\bibitem{WL11}
 \textsc{Z.\,F. Wang} and  \textsc{F.~Lui},
 \jr{Nanoscale} \textbf{3}, 4201 (2011).


\end{thebibliography}

\end{document}